\documentclass[conference]{IEEEtran}

\usepackage{amsmath, amssymb, graphicx, xcolor, multirow, subcaption, algorithm2e, url, float, booktabs, colortbl, comment}

\title{\textit{Avengers Assemble}: Amalgamation of Non-Semantic Features for Depression Detection}

\author{
    \IEEEauthorblockN{
        Orchid Chetia Phukan\IEEEauthorrefmark{1}\thanks{$^{\dagger}$Corresponding Author},
        Swarup Ranjan Behera\IEEEauthorrefmark{2},
        Shubham Singh\IEEEauthorrefmark{3}, 
        Muskaan Singh\IEEEauthorrefmark{4}\\
        Vandana Rajan\IEEEauthorrefmark{5}, 
        Arun Balaji Buduru\IEEEauthorrefmark{1}, 
        Rajesh Sharma\IEEEauthorrefmark{1}\IEEEauthorrefmark{6},
        S. R. Mahadeva Prasanna\IEEEauthorrefmark{7}\IEEEauthorrefmark{8}
    }
    \IEEEauthorblockA{
        \IEEEauthorrefmark{1}\textit{IIIT-Delhi, India},
        \IEEEauthorrefmark{2}\textit{Reliance Jio AICoE, India},
        \IEEEauthorrefmark{3}\textit{BIT Mesra, India},
        \IEEEauthorrefmark{4}\textit{ISRC, Ulster University, UK}\\
        \IEEEauthorrefmark{5}\textit{Independent Researcher, UK},
        \IEEEauthorrefmark{6}\textit{University of Tartu, Estonia},
        \IEEEauthorrefmark{7}\textit{IIT-Dharwad, India},
        \IEEEauthorrefmark{8}\textit{IIIT-Dharwad, India}\\
        \texttt{orchidp@iiitd.ac.in}
    }
}

\begin{document}

\maketitle
 
\begin{abstract}
In this study, we address the challenge of depression detection from speech, focusing on the potential of non-semantic features (NSFs) to capture subtle markers of depression. While prior research has leveraged various features for this task, NSFs-extracted from pre-trained models (PTMs) designed for non-semantic tasks such as paralinguistic speech processing (TRILLsson), speaker recognition (x-vector), and emotion recognition (emoHuBERT)-have shown significant promise. However, the potential of combining these diverse features has not been fully explored. In this work, we demonstrate that the amalgamation of NSFs results in complementary behavior, leading to enhanced depression detection performance. Furthermore, to our end, we introduce a simple novel framework, \textbf{FuSeR}, designed to effectively combine these features. Our results show that \textbf{FuSeR} outperforms models utilizing individual NSFs as well as baseline fusion techniques and obtains state-of-the-art (SOTA) performance in E-DAIC benchmark with RMSE of 5.51 and MAE of 4.48, establishing it as a robust approach for depression detection.
\end{abstract}

\vspace{-0.2cm}
\section{Introduction}
    Depression is the silent epidemic of the 21st century, affecting millions but spoken of by few. With over 300 million people worldwide suffering from depression, the condition not only diminishes quality of life but also contributes to severe economic and social burdens\footnote{https://www.who.int/news-room/fact-sheets/detail/depression}. The insidious nature of depression often leads to delayed diagnosis and treatment, exacerbating its effects. Early detection is critical, yet traditional methods rely heavily on self-reporting, which may not always be accurate or timely. As a result, there is an increasing need for innovative, objective methods to identify depression, particularly in its early stages. In this context, leveraging speech as a medium for detecting depressive symptoms offers a non-invasive and potentially more accessible approach to diagnosis.

    Initial efforts in speech-based depression detection centered on acoustic features such as low pitch, reduced pitch variability, slower speaking rate, increased pause frequency, and alterations in voice quality-markers indicative of depressive symptoms~\cite{alghowinem2013joyous, yang2012detecting, mundt2012vocal}. These paralinguistic features, encompassing prosody, voice quality, and articulation, were initially analyzed using classical ML algorithms like Gaussian Mixture Models (GMMs)~\cite{helfer2013classification, williamson2013vocal}, SVMs~\cite{cummins2013spectro, nasir2016multimodal}, and Logistic Regression (LR)~\cite{jan2017artificial}. The advent of deep learning brought a paradigm shift, enabling end-to-end neural networks such as CNNs and LSTMs to capture complex speech patterns more effectively~\cite{trigeorgis2016adieu, othmani2021towards}. Recent advancements have seen the introduction of attention-based mechanisms and multi-modal approaches, leveraging audio, video, and text data to enhance model robustness and accuracy~\cite{zhao2020hierarchical, yang2017multimodal}. Furthermore, with the utilization of pre-trained models (PTMs) designed for non-semantic tasks, including TRILLsson for paralinguistic speech processing~\cite{campbell23_interspeech}, x-vector for speaker recognition~\cite{egas2022automatic}, and models trained for speech emotion recognition (SER)~\cite{wu2022climate}, further enriched the feature sets employed in depression detection, yielding significant performance improvements in speech-based depression detection. These non-semantic features (NSFs) excel at capturing subtle vocal patterns indicative of depressive states, outperforming traditional semantic features. 

    Despite these advancements, the majority of research has concentrated on individual NSFs sets, leaving untapped potential in their combined use. 
    These fragmented focus on individual NSFs overlooks the opportunity to leverage their complementary strengths collectively. This raises a critical research question: \textit{How can we effectively amalgamate multiple NSFs to enhance depression detection performance, and what are the optimal methods for doing so?}

    To address this question, our research presents the following contributions:
    
    \begin{itemize} \item A comprehensive investigation into the fusion of various NSFs, including x-vector, TRILLsson, and from PTM trained for emotion recognition (emoHuBERT) and an evaluation of different fusion techniques. This work tackles the key question of \textit{``What to Fuse and How to Fuse?''} to maximize depression detection accuracy.

    \item A novel and simple feature fusion framework, \textbf{FuSeR}, which incorporates a bilinear pooling-based feature interaction for effective fusion. \textbf{FuSeR} with x-vector, TRILLsson, and emoHuBERT features demonstrates enhanced performance compared to models using individual features and those employing baseline fusion methods. With \textbf{FuSeR}, we also report SOTA results in the E-DAIC benchmark with RMSE of 5.51 and MAE of 4.48 in comparison to the previous SOTA.  \end{itemize}

    \noindent We will release the models and codes made for this study after the reviewing process. 


\section{Methodology}

In this section, we provide an overview of the NSFs utilized in our experiments,  baseline fusion techniques, and the proposed framework, \textbf{FuSeR}.

    \subsection{Feature Representations}

    \noindent \textbf{x-vector~\cite{8461375}:} It is a time-delay neural network trained for speaker recognition. It shows SOTA performance compared to its predecessor, i-vector. We include x-vector in our experiments as it has shown effectiveness in depression detection~\cite{egas2022automatic} and as well as related tasks like SER~\cite{Phukan2023TransformingTE}. For this study, we use the speechbrain x-vector model\footnote{\url{https://huggingface.co/speechbrain/spkrec-xvect-voxceleb}}. We extract features of 512-dimension by averaging across time. 
    
    \noindent \textbf{TRILLsson~\cite{shor22_interspeech}:} It is distilled from the SOTA universal paralinguistic conformer (CAP12) model that has shown SOTA in various non-semantic or paralinguistic tasks such as SER, speaker recognition, deepfake detection, and so on. TRILLsson is open-sourced, but CAP12 is not. TRILLsson also achieves near SOTA performance on the Non-Semantic Speech (NOSS) benchmark. For this study, we utilize the 1024-dimensional feature vectors generated from TRILLsson\footnote{\url{https://tfhub.dev/google/nonsemantic-speech-benchmark/trillsson4/1}}.
    
    \noindent \textbf{emoHuBERT:} For extracting emotion-specific features, we fine-tune HuBERT\footnote{\url{https://huggingface.co/facebook/hubert-base-ls960}}~\cite{hsu2021hubert}, originally pre-trained in a self-supervised manner for comprehensive speech tasks and name it as, emoHuBERT. We train it for 50 epochs on CREMA-D~\cite{cao2014crema}, a benchmark SER dataset, by attaching a probing head on top of the HuBERT model architecture. We unfreeze all the HuBERT layers for fine-tuning and extract features of 768 from the last hidden state by mean pooling for depression detection.  

    \begin{figure}[bt]
    \centering
    \includegraphics[scale=0.3]{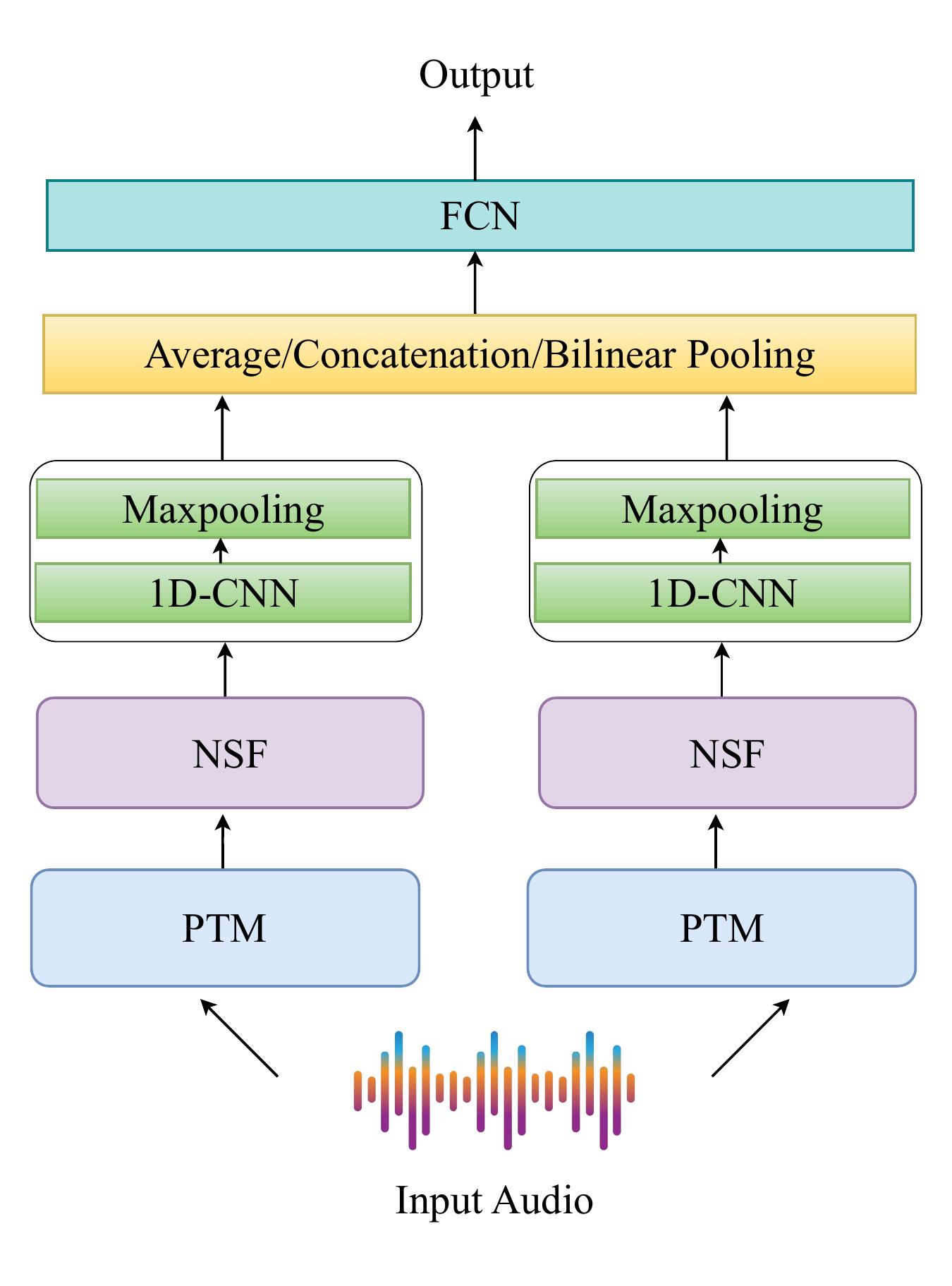}
    \caption{Modeling architecture for average/concatenation-based fusion and FuSeR; FuSeR leverages bilinear pooling as the fusion mechanism.}
    \label{fig:work_flow}
    \end{figure}
     \vspace{-0.2cm}
    \subsection{Fusion Techniques}
     \label{fusion}
    Here, we discuss various fusion techniques employed in our study for the fusion of NSFs and the downstream modeling employed with the fusion techniques. For a detailed visualization, refer to Figure \ref{fig:work_flow}.


    \noindent\textbf{Early Fusion:} It is a basic type of fusion, where we concat the NSFs directly after being extracted from the PTMs. We then use a convolutional block with a 1D-CNN layer of 32 filters with a $3\times3$ filter size, followed by a max-pooling layer. We then append a FCN module on top of the flattened features after maxpooling. For the FCN module, we attach a dense layer of 120 neurons followed by a output neuron that predicts a continous variable.

    \noindent\textbf{Average Fusion:} We add convolutional blocks with the same architectural details as used in Early Fusion CNN model on the top of each individual NSFs. We flatten the features after max-pooling followed by linear projection to the same dimension; we average the features from different NSFs and attach an FCN module on top of the averaged features  (See Figure \ref{fig:work_flow}). For FCN module, we use the same modeling details as Early Fusion FCN module. 
    
    \noindent\textbf{Concatenation Fusion:} Here, also we add convolutional block with the same architectural details as used in Early Fusion CNN model on the top of each individual NSFs. After flattening, we concat the features from different NSFs and attach a FCN module on top of the concatenated features  (See Figure \ref{fig:work_flow}). For the FCN module, we apply the same modeling details as in the Early Fusion FCN module.
       
    \vspace{-0.2cm}
    \subsection{Proposed Feature Fusion Framework - \textbf{FuSeR}}

    We introduce \textbf{FuSeR}, a simple yet effective feature fusion framework that makes use of bilinear pooling that as shown SOTA performance in multimodal tasks~\cite{kumar2022hate}. The proposed framework is shown in Figure~\ref{fig:work_flow}. First, we add convolutional blocks with the same architectural details as used in Early Fusion CNN model on the top of each individual NSFs. After flattening the features, we linear project it to 120 dimension. After this, we apply bilinear pooling that performs outer product on top of the projected features.  Mathematically, given two features \( \mathbf{a} \) and \( \mathbf{b} \), with dimensions \( C_A \times 1 \) and \( C_B \times 1 \) respectively, their outer product yields a matrix \( \mathbf{M} \) of dimensions \( C_A \times C_B \), represented as:
    \[
    \mathbf{M} = \mathbf{a} \otimes \mathbf{b}^T
    \]
    Here, \( \otimes \) denotes the outer product operation, and \( \mathbf{b}^T \) is the transpose of vector \( \mathbf{b} \). The matrix \( \mathbf{M} \) encapsulates the pairwise interactions between features from two different NSFs, resulting in a richer representation. The resulting matrix \( \mathbf{M} \) is flattened and passed to an FCN module with similar modeling details to the Early Fusion FCN module, followed by an output neuron.

\vspace{-0.4cm}
    \section{Experiments}
    \subsection{Benchmark Dataset}
    We employ the Extended DAIC (E-DAIC) dataset~\cite{gratch2014distress, ringeval2019avec}, an enhanced version of the DAIC-WOZ dataset from the AVEC 2019 challenge, which includes semi-structured clinical information consistent with DAIC-WOZ. The E-DAIC dataset comprises data from 275 participants: 163 for training, 56 for development, and 56 for testing, specifically curated to aid in depression diagnosis. This extensive dataset supports robust model evaluation facilitating the development and validation of more effective depression detection methodologies.
    \vspace{-0.2cm}
    \subsection{Data Pre-Processing}
    We remove the silence and trim the audio to 5 seconds. Further, we resample all the audios to 16kHz before passing it to the PTMs for feature extraction. 
    \vspace{-0.2cm}
    \subsection{Training Details}
    We use Mean Squared Error (MSE) as the loss function and Adam as the optimizer. We employ a learning rate of 1e-3 and train all the models for 30 epochs, with a batch size of 64. We use the official split given in the dataset. We trained the models on training set, validated on validation set and tested on testing set. To evaluate model performance comprehensively, we utilize mean absolute error (MAE) and root mean squared error (RMSE) as the evaluation metrics keeping in line with previous research~\cite{ringeval2019avec, li23d_interspeech}. The training parameters for downstream models with different fusion techniques range between 0.2M to 1.1M. 

    \begin{table}[bt]
    \centering
    
    \caption{Evaluation results with downstream models trained on different NSFs individually for the E-DAIC dataset. For RMSE and MAE, lower scores indicate better performance.}

    \begin{tabular}{|l|l|l|l}
    \toprule
    \multicolumn{1}{c|}{\textbf{Feature}} & \multicolumn{1}{c|}{\textbf{Methods}} & \multicolumn{1}{c|}{\textbf{RMSE}} & {\textbf{MAE}} \\ \midrule
    
    \multicolumn{1}{c|}{\multirow{4}{*}{x-vector}} & SVR & 7.35  & 6.04\\ 
    \multicolumn{1}{c|}{} & RF & 6.96 & 5.91\\ 
    \multicolumn{1}{c|}{} & FCN & 6.93 & 5.35 \\
    \multicolumn{1}{c|}{} & CNN & 6.78 & 5.10 \\
    \midrule
    
    \multicolumn{1}{c|}{\multirow{4}{*}{emoHuBERT}} & SVR & 6.78 & 5.12 \\ 
    \multicolumn{1}{c|}{} & RF & 6.79 & 5.09 \\ 
    \multicolumn{1}{c|}{} & FCN & 6.78 & 5.07 \\
    \multicolumn{1}{c|}{} & CNN &  \cellcolor{yellow!45}\textbf{6.63} &  \cellcolor{yellow!45}\textbf{5.05}\\
    \midrule
    
    \multicolumn{1}{c|}{\multirow{4}{*}{TRILLsson}} & SVR & 6.72 &  \cellcolor{green!25}\textbf{5.06}  \\ 
    \multicolumn{1}{c|}{} & RF &  \cellcolor{green!25}\textbf{6.71} &  \cellcolor{yellow!45}\textbf{5.05} \\ 
    \multicolumn{1}{c|}{} & FCN & \cellcolor{blue!25}\textbf{6.56} &  \cellcolor{yellow!45}\textbf{5.05} \\
    \multicolumn{1}{c|}{} & CNN & \cellcolor{blue!25}\textbf{6.56} & \cellcolor{blue!25}\textbf{5.03}\\
    \midrule
    
    \end{tabular}\label{e_single}
    \end{table}

    \begin{table}[bt]
    \centering
    \caption{Evaluation results with downstream models trained on the fusion of different NSFs for the E-DAIC dataset. For RMSE and MAE, lower scores indicate better model performance.}
    \begin{tabular}{|l|l|l|l}
    \toprule
    \multicolumn{1}{c|}{\textbf{Feature}} & \multicolumn{1}{c|}{\textbf{Fusion}} & \multicolumn{1}{c|}{\textbf{RMSE}} & {\textbf{MAE}} \\ \midrule
    
    \multicolumn{1}{c|}{\multirow{4}{*}{\shortstack{x-vector \\ + \\ emoHuBERT}}} & Early & 6.63 &  5.02\\ 
    \multicolumn{1}{c|}{} & Average & 6.56 &  5.02\\ 
    \multicolumn{1}{c|}{} & Concat & 6.55 & 4.99\\
    \multicolumn{1}{c|}{} & \textbf{FuSeR} & 6.43 & 4.98\\
    \midrule
    
    \multicolumn{1}{c|}{\multirow{4}{*}{\shortstack{x-vector \\ + \\ TRILLsson}}} & Early & 6.48 & 5.01\\ 
    \multicolumn{1}{c|}{} & Average & 6.40 & 4.95\\ 
    \multicolumn{1}{c|}{} & Concat & 6.40 & 4.96 \\
    \multicolumn{1}{c|}{} & \textbf{FuSeR} & 6.32 & 4.95  \\
    \midrule
    
    \multicolumn{1}{c|}{\multirow{4}{*}{\shortstack{emoHuBERT \\ + \\ TRILLsson}}} & Early & 6.28 & 4.92\\ 
    \multicolumn{1}{c|}{} & Average & 6.24 &  4.92 \\ 
    \multicolumn{1}{c|}{} & Concat &  6.27 & 4.84  \\
    \multicolumn{1}{c|}{} & \textbf{FuSeR} & \cellcolor{green!25}\textbf{6.06} & \cellcolor{green!25}\textbf{4.79}\\
    \midrule
    
    \multicolumn{1}{c|}{\multirow{4}{*}{\shortstack{x-vector \\+\\ emoHuBERT \\ + \\ TRILLsson}}} & Early & 6.24 &4.90 \\ 
    \multicolumn{1}{c|}{} & Average & 6.16 & 4.80 \\ 
    \multicolumn{1}{c|}{} & Concat &\cellcolor{yellow!45}\textbf{6.00}& \cellcolor{yellow!45}\textbf{4.71}\\
    \multicolumn{1}{c|}{} & \textbf{FuSeR} & \cellcolor{blue!25}\textbf{5.51} & \cellcolor{blue!25}\textbf{4.48}\\
    \midrule
    \end{tabular}\label{e_combined}
    \end{table}
    
    \begin{table}[t]
    \centering
    \caption{Comparison to SOTA on the E-DAIC dataset; xv, eH, and TL represent x-vector, emoHuBERT, and TRILLsson, respectively.}
    \begin{tabular}{l|l|l|l}
    \toprule
    \textbf{Feature} & \textbf{Method} & \textbf{RMSE} & \textbf{MAE} \\ \midrule
    
    DS-VGG~\cite{ringeval2019avec} & GRU~\cite{ringeval2019avec} & 9.33 &  - \\

    BoAW-e~\cite{ringeval2019avec} & GRU~\cite{ringeval2019avec} & 8.19 &  - \\
    
    xv + eH +TL & \textbf{FuSeR} & \cellcolor{blue!25}\textbf{5.51} & \cellcolor{blue!25}\textbf{4.48}  \\
    \midrule
    \end{tabular}\label{e_sota}
    \end{table}

\vspace{-0.2cm}
\subsection{Experimental Results}
Firstly, as baseline experiments, we train downstream models on top of the three NSFs considered in our study. We train different downstream models, such as Support Vector Regression (SVR), Random Forest (RF), Fully Connected Network (FCN), and lastly CNN. We use the default parameters as given in \textit{Scikit-learn} library for SVR and RF. For FCN, we have two dense layers of 120 and 90 neurons, followed by the output that predicts a continuous variable. We follow the same modeling for the CNN model as followed in Section \ref{fusion} Early Fusion technique. We follow the same training and evaluation regime as the fusion techniques. \par

The scores for all the tables are scores on the test set. Table~\ref{e_single} details the performance of different downstream models trained on various NSFs on E-DAIC. Among the NSFs evaluated, TRILLsson with CNN, achieves the lowest RMSE (6.56) and MAE (5.03) scores, indicating superior performance in capturing subtle markers of depression. Conversely, the x-vector features exhibit comparatively higher error rates than the other NSFs. 

Table~\ref{e_combined} presents the evaluation results of downstream models trained on different combinations of NSFs, employing various fusion techniques. The results demonstrate that \textbf{FuSeR}, the proposed fusion framework, consistently outperforms the baseline fusion techniques such as Early, Average, and Concatenation methods as well as the individual NSFs. Specifically, when combining x-vector and emoHuBERT, \textbf{FuSeR} achieves an RMSE of 6.43 and an MAE of 4.98, outperforming the concatenated fusion approach with an RMSE of 6.55 and MAE of 4.99. Notably, the combination of all three NSFs - x-vector, emoHuBERT, and TRILLsson - with \textbf{FuSeR} yields the topmost performance delivering an RMSE of 5.51 and an MAE of 4.48. This highlights the effectiveness of \textbf{FuSeR} in integrating diverse feature sets to enhance model performance.

Table~\ref{e_sota} provides a comparison of \textbf{FuSeR} approach against SOTA on the E-DAIC dataset. The results reveal that \textbf{FuSeR}, using the combination of x-vector, emoHuBERT, and TRILLsson, achieves the lowest RMSE of 5.51 and MAE of 4.48, outperforming previous SOTA work. For instance, the GRU model with DS-VGG and BoAW-e features achieves higher RMSE values of 9.33 and 8.19, respectively, demonstrating that \textbf{FuSeR} not only improves upon current methods but sets a new benchmark in depression detection performance.


\vspace{-0.2cm}
\subsection{Additional Experiments}
    In our additional experiments, we evaluated the generalizability of the proposed FuSeR framework on the ANDROID dataset~\cite{tao23_interspeech}, a recent Italian dataset designed for depression detection. This dataset includes tasks such as Interview (I) and Reading (R). Notably, the proposed framework, \textbf{FuSeR} also demonstrates strong performance in a binary depression detection (Yes/No) scenario, highlighting its versatility and robustness across different types of tasks. For all the experiments on ANDROID, we follow the same pre-processing steps, modeling, and training details as followed in E-DAIC.

    Table~\ref{an_single} presents the results for downstream models trained on individual NSFs for the ANDROID dataset. Among the NSFs evaluated, TRILLsson consistently demonstrates superior performance across all models. Specifically, CNN with TRILLsson achieves the highest Accuracy and F1-score for both the Interview (80.52\% and 78.68\%) and Reading (75.13\% and 73.94\%) tasks, indicating that TRILLsson features provide the most robust representation for depression detection. In contrast, x-vector features show relatively lower performance, with CNN achieving 66.32\% Accuracy and 64.71\% F1-score for the Interview task, and 74.11\% Accuracy and 70.12\% F1-score for the Reading task.

    Table~\ref{an_combined} showcases the performance of various feature fusion techniques on the ANDROID dataset. The proposed \textbf{FuSeR} framework significantly outperforms other fusion methods such as Early, Average, and Concatenation across all feature combinations. For example, with the combination of x-vector and emoHuBERT features, \textbf{FuSeR} achieves an Accuracy of 68.91\% and an F1-score of 67.73\% for the Interview task, and 75.38\% Accuracy and 75.31\% F1-score for the Reading task, surpassing the concatenated approach (65.39\% and 64.29\% for the Interview task, and 75.13\% and 72.19\% for the Reading task).

    When combining x-vector, emoHuBERT, and TRILLsson through \textbf{FuSeR}, we achieve the topmost results, with an Accuracy of 87.93\% and an F1-score of 87.84\% for the Interview task, and 84.72\% Accuracy and 83.35\% F1-score for the Reading task. These results underscore \textbf{FuSeR}’s capacity to leverage complementary information from different NSFs effectively and setting a new SOTA on ANDROID dataset. Overall, these additional experiments validate the robustness and adaptability of the \textbf{FuSeR} for improved depression detection.

\begin{table}[bt]
\centering
\caption{Evaluation results with models trained on different NSFs for ANDROID. Acc(I), F1(I), Acc(R), and F1(R) represent accuracy and macro-average F1-score for the Interview (I) and Reading (R) tasks, respectively. Scores are in \%.}
\begin{tabular}{|l|l|l|l|l|l}
\toprule
\multicolumn{1}{c|}{\textbf{Feature}} & \multicolumn{1}{c|}{\textbf{Methods}} & \multicolumn{1}{c|}{\textbf{Acc(I)}} & {\textbf{F1(I)}} & {\textbf{Acc(R)}} & {\textbf{F1(R)}} \\ \midrule 

\multicolumn{1}{c|}{\multirow{4}{*}{x-vector}} & SVM & 62.49 & 60.49 & 60.91 & 60.81 \\ 
\multicolumn{1}{c|}{} & RF &  63.83 & 61.20 & 59.90 & 59.48 \\ 
\multicolumn{1}{c|}{} & FCN & 64.97 & 61.70 & 64.97 & 64.61\\
\multicolumn{1}{c|}{} & CNN & 66.32 & 64.71 & \cellcolor{yellow!45}\textbf{74.11} & 70.12\\
\midrule

\multicolumn{1}{c|}{\multirow{4}{*}{emoHuBERT}} & SVM & 64.25 & 61.16 & 61.42 & 69.09 \\ 
\multicolumn{1}{c|}{} & RF & 73.89 & 71.86 & 71.06 & 69.07 \\ 
\multicolumn{1}{c|}{} & FCN & 75.65 & 72.39& 71.57 & 69.17 \\
\multicolumn{1}{c|}{} & CNN & 75.96 & 73.68 & 71.57 & 70.29\\
\midrule

\multicolumn{1}{c|}{\multirow{4}{*}{TRILLsson}} & SVM & 69.02 & 64.91  & 64.97 & 64.03\\ 
\multicolumn{1}{c|}{} & RF & \cellcolor{yellow!45}\textbf{79.38} & \cellcolor{green!25}\textbf{77.42} & 68.02  & 67.36\\ 
\multicolumn{1}{c|}{} & FCN &  \cellcolor{yellow!45}\textbf{79.38} & \cellcolor{yellow!45}\textbf{77.57} &  \cellcolor{green!25}\textbf{72.59} & \cellcolor{yellow!45}\textbf{71.69}\\
\multicolumn{1}{c|}{} & CNN & \cellcolor{blue!25}\textbf{80.52} & \cellcolor{blue!25}\textbf{78.68} & \cellcolor{blue!25}\textbf{75.13} & \cellcolor{blue!25}\textbf{73.94}\\
\midrule

\end{tabular}\label{an_single}
\end{table}

\begin{table}[bt]
\centering
\caption{Evaluation results with models trained on the combination of different NSFs for ANDROID. Acc(I), F1(I), Acc(R), and F1(R) represent accuracy and F1-score for Interview (I) and Reading (R) tasks, respectively. Scores are in \%.}
\begin{tabular}{|l|l|l|l|l|l}
\toprule
\multicolumn{1}{c|}{\textbf{Feature}} & \multicolumn{1}{c|}{\textbf{Fusion}} & \multicolumn{1}{c|}{\textbf{Acc(I)}} & {\textbf{F1(I)}} & {\textbf{Acc(R)}} & {\textbf{F1(R)}} \\ \midrule 

\multicolumn{1}{c|}{\multirow{4}{*}{\shortstack{x-vector \\ + \\ emoHuBERT}}} & Early & 63.52 & 62.16 & 71.57 & 69.17 \\ 
\multicolumn{1}{c|}{} & Average & 66.32 & 64.78 & 72.08 & 69.23 \\ 
\multicolumn{1}{c|}{} & Concat & 65.39 &  64.29& 75.13 & 72.19\\
\multicolumn{1}{c|}{} & \textbf{FuSeR} & 68.91 & 67.73 & 75.38 & 75.31\\
\midrule

\multicolumn{1}{c|}{\multirow{4}{*}{\shortstack{x-vector \\ + \\ TRILLsson}}} & Early & 69.12 & 68.89 & 68.53 & 67.38 \\ 
\multicolumn{1}{c|}{} & Average & 73.47 & 70.78 & 72.59 & 69.00\\ 
\multicolumn{1}{c|}{} & Concat &  74.82 & 73.69 & 78.17 & 75.60  \\
\multicolumn{1}{c|}{} & \textbf{FuSeR} & 80.41 & 78.78 & 78.17 & 76.67 \\
\midrule

\multicolumn{1}{c|}{\multirow{4}{*}{\shortstack{emoHuBERT \\ + \\ TRILLsson}}} & Early & 74.30 & 72.88 & 71.07& 70.38\\ 
\multicolumn{1}{c|}{} & Average &  74.72 & 73.03 & 71.57 & 70.01 \\ 
\multicolumn{1}{c|}{} & Concat & 76.06 & 74.51 & 73.10 & 72.17\\
\multicolumn{1}{c|}{} & \textbf{FuSeR} & \cellcolor{yellow!45}\textbf{83.32} & \cellcolor{yellow!45}\textbf{81.73} & \cellcolor{yellow!45}\textbf{80.20} & \cellcolor{yellow!45}\textbf{77.34}\\
\midrule

\multicolumn{1}{c|}{\multirow{4}{*}{\shortstack{x-vector \\ + \\ emoHuBERT \\ + \\ TRILLsson}}} & Early & 78.45 & 76.60 & 72.59 & 71.70\\ 
\multicolumn{1}{c|}{} & Average & \cellcolor{green!25}\textbf{79.38} & \cellcolor{green!25}\textbf{77.63} & 74.62 & 72.95\\ 
\multicolumn{1}{c|}{} & Concat & 78.76 & 77.41 & \cellcolor{green!25}\textbf{76.65} & \cellcolor{green!25}\textbf{74.35}\\
\multicolumn{1}{c|}{} & \textbf{FuSeR} & \cellcolor{blue!25}\textbf{87.93} & \cellcolor{blue!25}\textbf{87.84} & \cellcolor{blue!25}\textbf{84.72} & \cellcolor{blue!25}\textbf{83.35}\\
\midrule

\end{tabular}\label{an_combined}
\end{table}

\vspace{-0.4cm}
\section{Conclusion}
In this work, we explore and show the potential of exploiting the complementary strength of NSFs to capture subtle markers of depression for better depression detection. We explore combination of different NSFs such as from PTMs trained for paralinguistic speech processing (TRILLs-son), speaker recognition (x-vector), and emotion recognition (emoHuBERT) using different fusion strategies. We also propose a simple novel framework, \textbf{FuSeR} for effective fusion of the NSFs. \textbf{FuSeR} outperforms individual NSFs and baseline fusion methods such as early fusion, average fusion, and concatenation fusion, achieving SOTA results on the E-DAIC benchmark with an RMSE of 5.51 and an MAE of 4.48, demonstrating its robustness for depression detection. Our findings demonstrate that the fusion of these diverse NSFs results in complementary behavior, significantly enhancing depression detection performance. Our study will inspire future research in coming up with more optimal fusion strategies for combining NSFs for depression detection and serve as a benchmark for the same. 

\bibliographystyle{IEEEtran}
\bibliography{mybib}

\begin{thebibliography}{10}
\providecommand{\url}[1]{#1}
\csname url@samestyle\endcsname
\providecommand{\newblock}{\relax}
\providecommand{\bibinfo}[2]{#2}
\providecommand{\BIBentrySTDinterwordspacing}{\spaceskip=0pt\relax}
\providecommand{\BIBentryALTinterwordstretchfactor}{4}
\providecommand{\BIBentryALTinterwordspacing}{\spaceskip=\fontdimen2\font plus
\BIBentryALTinterwordstretchfactor\fontdimen3\font minus \fontdimen4\font\relax}
\providecommand{\BIBforeignlanguage}[2]{{%
\expandafter\ifx\csname l@#1\endcsname\relax
\typeout{** WARNING: IEEEtran.bst: No hyphenation pattern has been}%
\typeout{** loaded for the language `#1'. Using the pattern for}%
\typeout{** the default language instead.}%
\else
\language=\csname l@#1\endcsname
\fi
#2}}
\providecommand{\BIBdecl}{\relax}
\BIBdecl

\bibitem{alghowinem2013joyous}
S.~Alghowinem, ``From joyous to clinically depressed: Mood detection using multimodal analysis of a person's appearance and speech,'' in \emph{2013 Humaine Association Conference on Affective Computing and Intelligent Interaction}.\hskip 1em plus 0.5em minus 0.4em\relax IEEE, 2013, pp. 648--654.

\bibitem{yang2012detecting}
Y.~Yang, C.~Fairbairn, and J.~F. Cohn, ``Detecting depression severity from vocal prosody,'' \emph{IEEE transactions on affective computing}, vol.~4, no.~2, pp. 142--150, 2012.

\bibitem{mundt2012vocal}
J.~C. Mundt, A.~P. Vogel, D.~E. Feltner, and W.~R. Lenderking, ``Vocal acoustic biomarkers of depression severity and treatment response,'' \emph{Biological psychiatry}, vol.~72, no.~7, pp. 580--587, 2012.

\bibitem{helfer2013classification}
B.~S. Helfer, T.~F. Quatieri, J.~R. Williamson, D.~D. Mehta, R.~Horwitz, and B.~Yu, ``Classification of depression state based on articulatory precision.'' in \emph{Interspeech}, 2013, pp. 2172--2176.

\bibitem{williamson2013vocal}
J.~R. Williamson, T.~F. Quatieri, B.~S. Helfer, R.~Horwitz, B.~Yu, and D.~D. Mehta, ``Vocal biomarkers of depression based on motor incoordination,'' in \emph{Proceedings of the 3rd ACM international workshop on Audio/visual emotion challenge}, 2013, pp. 41--48.

\bibitem{cummins2013spectro}
N.~Cummins, J.~Epps, and E.~Ambikairajah, ``Spectro-temporal analysis of speech affected by depression and psychomotor retardation,'' in \emph{2013 IEEE international conference on acoustics, speech and signal processing}.\hskip 1em plus 0.5em minus 0.4em\relax IEEE, 2013, pp. 7542--7546.

\bibitem{nasir2016multimodal}
M.~Nasir, A.~Jati, P.~G. Shivakumar, S.~Nallan~Chakravarthula, and P.~Georgiou, ``Multimodal and multiresolution depression detection from speech and facial landmark features,'' in \emph{Proceedings of the 6th international workshop on audio/visual emotion challenge}, 2016, pp. 43--50.

\bibitem{jan2017artificial}
A.~Jan, H.~Meng, Y.~F. B.~A. Gaus, and F.~Zhang, ``Artificial intelligent system for automatic depression level analysis through visual and vocal expressions,'' \emph{IEEE Transactions on Cognitive and Developmental Systems}, vol.~10, no.~3, pp. 668--680, 2017.

\bibitem{trigeorgis2016adieu}
G.~Trigeorgis, F.~Ringeval, R.~Brueckner, E.~Marchi, M.~A. Nicolaou, B.~Schuller, and S.~Zafeiriou, ``Adieu features? end-to-end speech emotion recognition using a deep convolutional recurrent network,'' in \emph{2016 IEEE international conference on acoustics, speech and signal processing (ICASSP)}.\hskip 1em plus 0.5em minus 0.4em\relax IEEE, 2016, pp. 5200--5204.

\bibitem{othmani2021towards}
A.~Othmani, D.~Kadoch, K.~Bentounes, E.~Rejaibi, R.~Alfred, and A.~Hadid, ``Towards robust deep neural networks for affect and depression recognition from speech,'' in \emph{Pattern Recognition. ICPR International Workshops and Challenges: Virtual Event, January 10--15, 2021, Proceedings, Part II}.\hskip 1em plus 0.5em minus 0.4em\relax Springer, 2021, pp. 5--19.

\bibitem{zhao2020hierarchical}
Z.~Zhao, Z.~Bao, Z.~Zhang, N.~Cummins, H.~Wang, and B.~Schuller, ``Hierarchical attention transfer networks for depression assessment from speech,'' in \emph{ICASSP 2020-2020 IEEE international conference on acoustics, speech and signal processing (ICASSP)}.\hskip 1em plus 0.5em minus 0.4em\relax IEEE, 2020, pp. 7159--7163.

\bibitem{yang2017multimodal}
L.~Yang, D.~Jiang, X.~Xia, E.~Pei, M.~C. Oveneke, and H.~Sahli, ``Multimodal measurement of depression using deep learning models,'' in \emph{Proceedings of the 7th Annual Workshop on Audio/Visual Emotion Challenge}, 2017, pp. 53--59.

\bibitem{campbell23_interspeech}
E.~L. Campbell, J.~Dineley, P.~Conde, F.~Matcham, K.~M. White, C.~Oetzmann, S.~Simblett, S.~Bruce, A.~A. Folarin, T.~Wykes, S.~Vairavan, R.~J.~B. Dobson, L.~Docio-Fernandez, C.~Garcia-Mateo, V.~A. Narayan, M.~Hotopf, and N.~Cummins, ``Classifying depression symptom severity: Assessment of speech representations in personalized and generalized machine learning models.'' in \emph{INTERSPEECH 2023}, 2023, pp. 1738--1742.

\bibitem{egas2022automatic}
J.~V. Egas-L{\'o}pez, G.~Kiss, D.~Sztah{\'o}, and G.~Gosztolya, ``Automatic assessment of the degree of clinical depression from speech using x-vectors,'' in \emph{ICASSP 2022-2022 IEEE International Conference on Acoustics, Speech and Signal Processing (ICASSP)}.\hskip 1em plus 0.5em minus 0.4em\relax IEEE, 2022, pp. 8502--8506.

\bibitem{wu2022climate}
W.~Wu, M.~Wu, and K.~Yu, ``Climate and weather: Inspecting depression detection via emotion recognition,'' in \emph{ICASSP 2022-2022 IEEE International Conference on Acoustics, Speech and Signal Processing (ICASSP)}.\hskip 1em plus 0.5em minus 0.4em\relax IEEE, 2022, pp. 6262--6266.

\bibitem{8461375}
D.~Snyder, D.~Garcia-Romero, G.~Sell, D.~Povey, and S.~Khudanpur, ``X-vectors: Robust dnn embeddings for speaker recognition,'' in \emph{2018 IEEE International Conference on Acoustics, Speech and Signal Processing (ICASSP)}, 2018, pp. 5329--5333.

\bibitem{Phukan2023TransformingTE}
O.~C. Phukan, A.~B. Buduru, and R.~Sharma, ``Transforming the embeddings: A lightweight technique for speech emotion recognition tasks,'' in \emph{Interspeech}, 2023.

\bibitem{shor22_interspeech}
J.~Shor and S.~Venugopalan, ``{TRILLsson: Distilled Universal Paralinguistic Speech Representations},'' in \emph{Proc. Interspeech 2022}, 2022, pp. 356--360.

\bibitem{hsu2021hubert}
W.-N. Hsu, B.~Bolte, Y.-H.~H. Tsai, K.~Lakhotia, R.~Salakhutdinov, and A.~Mohamed, ``Hubert: Self-supervised speech representation learning by masked prediction of hidden units,'' \emph{IEEE/ACM Transactions on Audio, Speech, and Language Processing}, vol.~29, pp. 3451--3460, 2021.

\bibitem{cao2014crema}
H.~Cao, D.~G. Cooper, M.~K. Keutmann, R.~C. Gur, A.~Nenkova, and R.~Verma, ``Crema-d: Crowd-sourced emotional multimodal actors dataset,'' \emph{IEEE transactions on affective computing}, vol.~5, no.~4, pp. 377--390, 2014.

\bibitem{kumar2022hate}
G.~K. Kumar and K.~Nandakumar, ``Hate-clipper: Multimodal hateful meme classification based on cross-modal interaction of clip features,'' \emph{arXiv preprint arXiv:2210.05916}, 2022.

\bibitem{gratch2014distress}
J.~Gratch, R.~Artstein, G.~M. Lucas, G.~Stratou, S.~Scherer, A.~Nazarian, R.~Wood, J.~Boberg, D.~DeVault, S.~Marsella \emph{et~al.}, ``The distress analysis interview corpus of human and computer interviews.'' in \emph{LREC}.\hskip 1em plus 0.5em minus 0.4em\relax Reykjavik, 2014, pp. 3123--3128.

\bibitem{ringeval2019avec}
F.~Ringeval, B.~Schuller, M.~Valstar, N.~Cummins, R.~Cowie, L.~Tavabi, M.~Schmitt, S.~Alisamir, S.~Amiriparian, E.-M. Messner \emph{et~al.}, ``Avec 2019 workshop and challenge: state-of-mind, detecting depression with ai, and cross-cultural affect recognition,'' in \emph{Proceedings of the 9th International on Audio/visual Emotion Challenge and Workshop}, 2019, pp. 3--12.

\bibitem{li23d_interspeech}
Q.~Li, D.~Wang, Y.~Ren, Y.~Gao, and Y.~Li, ``Fta-net: A frequency and time attention network for speech depression detection,'' in \emph{INTERSPEECH 2023}, 2023, pp. 1723--1727.

\bibitem{tao23_interspeech}
F.~Tao, A.~Esposito, and A.~Vinciarelli, ``{The Androids Corpus: A New Publicly Available Benchmark for Speech Based Depression Detection},'' in \emph{Proc. INTERSPEECH 2023}, 2023, pp. 4149--4153.

\end{thebibliography}

\end{document}